\documentclass[aps,floatfix,showpacs,showkeys,amsmath]{revtex4-1}
\usepackage{color,graphicx}
\usepackage{dcolumn}
\usepackage{bm}
\usepackage{epsfig}
\usepackage{supertabular}
\usepackage[bookmarksopen]{hyperref}
\def\be {\begin{equation}}
\def\ee {\end{equation}}
\def\nn {\nonumber}
\def\bea {\begin{eqnarray}}
\def\eea {\end{eqnarray}}

\def  \bef  {\begin{figure}}
\def  \eef  {\end{figure}}
\def  \be   {\begin{equation}}
\def  \ee   {\end{equation}}
\def  \ba   {\begin{array}}
\def  \ea   {\end{array}}
\def  \bea  {\begin{eqnarray}}
\def  \eea  {\end{eqnarray}}
\def  \beq  {\begin{eqnarray}}
\def  \eeq  {\end{eqnarray}}
\def  \nn   {\nonumber}
\def  \bd   {\begin{displaymath}}
\def  \ed   {\end{displaymath}}
\def  \bse  {\begin{subequations}}
\def  \ese  {\end{subequations}}
\def  \bwt  {\begin{widetext}}
\def  \ewt  {\end{widetext}}

\def  \ba   {{\bf{a_1}}}

\begin{document}
	\title{Pion mass modification in presence of external magnetic field}
	%
	%
	\author{S. P. Adhya$^1$, M. Mandal$^2$, S. Biswas$^3$ and P. K. Roy$^4$}
	%
	%
	%
	\affiliation{$1$ Experimental High Energy Physics and Applications Group, Variable Energy Cyclotron Cente, 1/AF Bidhannagar, Kolkata-700 064, INDIA\\
		$^2$Department of Physics, Government General Degree College, Kalna-I, Burdwan-713405, INDIA\\
		$^3$ Department of Physics, Rishi Bankimchandra College, Naihati, Pin-743165, INDIA\\
		$^4$High Energy Nuclear and Particle Physics Division, Saha Institute
		of Nuclear Physics,
		1/AF Bidhannagar, Kolkata-700 064, INDIA\\
		\email{sp.adhya@vecc.gov.in}
	}
	\begin{abstract}
		In this work, the self energies of $\pi_0$ and $\pi_{\pm}$ up to one loop order have been calculated in the limit of weak
		external magnetic field. The effective
		masses are explicitly dependent on the magnetic field which are modified significantly for the pseudoscalar coupling due to weak field approximation of the external field. On the other hand, for the pseudovector coupling, there is a modest increment in the effective masses of the pions. These theoretical developments are relevant for the study of the phenomenological aspect of mesons in the context of neutron stars as well as heavy ion collisions.
	\end{abstract}
	\maketitle
	\section{Introduction}
	The modifications due to the magnetic field on various theoretical estimates in heavy ion collisions as well as neutron star have drawn significant attention in recent years \cite{Tuchin:2013bda,Ayala:2015lta,Colucci:2013zoa,Ghosh:2016evc}. As for example, in heavy ion collisions for off central collisions, magnetic field can be $\sim 0.02$ $GeV^2$ (at RHIC) and $\sim 0.3$ $GeV^2$ (at LHC). On the other hand, field strengths of $\sim 10^{15}$ Gauss can be found in some neutron stars. Thus, it will be worthwhile to calculate the mass modification of pions in such scenarios because of the comparable mass and field strengths. For our purpose, we use the Schwinger's proper time approach for calculation of the magnetic field modified Fermionic propagators \cite{schwinger51}. A consistent calculation of the pion self energy has been presented for both the pseudo-scalar and pseudo-vector coupling of the pion-nucleon Lagrangian. However, we present the results in absence of any medium. The modification of the dispersion relations in presence of the medium and magnetic field will be reported shortly. 
	\section{Formalism}
	\begin{figure}[htb]
		\begin{center}
			\includegraphics[scale=0.5,angle=0]{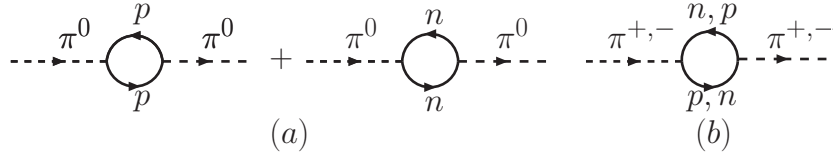}
			\caption{(a) represents the one-loop self-energy diagram for $\pi^0$
				and (b) represents the same for $\pi^{\pm}$.}
			\label{fig0}
		\end{center}
	\end{figure}
	
	The pion-nucleon phenomenological Lagrangian is written as,
	\bea
	{\mathcal L}_{\rm int}^{\rm PS} =  -i{\rm g}_{\pi}{\bar \Psi}
	\gamma_5({\vec \tau}\cdot {\vec \Phi_{\pi}})\Psi~~~~~~~~~(pseudo-scalar~coupling)\label{lag1}\\
	\mathcal{L}_{\rm int}^{\rm PV} = -\frac{f_{\pi}}{m_{\pi}}{\bar\Psi}^{\prime}\gamma_5\gamma^{\mu}
	\partial_{\mu}({\bf{\tau}} .\Phi_{\pi}^{\prime})\Psi^{\prime}~~~~~~~~~(pseudo-vector~coupling)
	\eea
	where ${\mathcal L}_{\rm int}^{\rm PS}$ and $\mathcal{L}_{\rm int}^{\rm PV}$ are the Lagrangians corresponding pseudo-scalar and pseudo-vector couplings respectively. In the above equation, $(\Psi,\Psi^{\prime})$ and $(\Phi,\Phi^{\prime})$ are the nucleon and pion fields with the coupling constants $g_{\pi}$ and $f_{\pi}$ accordingly.
	Therefore, the pion self energy at one loop order is given as,
	\be
	\Pi_{\pi}(q) = -i\int \frac{d^4k}{(2\pi)^4}{\rm Tr}[\{i\Gamma(q)\}iS_a(k)\{i\Gamma(-q)\}iS_b(k+q)]
	\ee
	We consider the weak field approximation of the magnetic field on the propagators according to the condition $(eB<< m_\pi^2)$ .Thus up to order $(eB)^2$, the Fermion propagators can be written as \cite{adhya17},
	\be
	S(k) = S^{(0)}(k)+eB\, S^{(1)}(k) + (eB)^2\, S^{(2)}(k) + {\mathcal O}((eB)^3) 
	\ee
	The form of $S^{(1)}(k)$ and $S^{(2)}(k)$ can be found in the Ref.\cite{adhya17}.
	Thus, using the modified propagator, we arrive at \cite{adhya17},
	\bea
	&&\Pi_{\pi^0}(q)^{PS} = -\frac{g^2_{\pi}}{2\pi^2}\Bigg[\int^1_0dx \Big[
	\frac{(q^2-m_{\pi^0}^2)\,x(1-x)(m^2-3q^2x(1-x))}{\Delta_R}\nn\\
	&+&(m^2-3m_{\pi^0}^2\,x(1-x))
	\log\frac{\Delta_R}{m^2-m_{\pi^0}^2\,x(1-x)}\Big]\nn\\
	&+&\frac{(eB)^2}{2}\int^1_0dx\,x(1-x)\Big(\frac{1}{\Delta_R}+
	\frac{m^2+x(1-x)q^2_{||}}{\Delta_R^2}\Big)\nn\\
	&+& (eB)^2\Bigg\{\int_0^1 dx\,(1-x)^3\big[\frac{1}{\Delta_R}+
	\frac{q^2 x(1-x)+q^2_{\bot}x(4x-1)+m^2}{3\Delta_R^2}\nn\\
	&+&\frac{2x^2q_{\bot}^2[q^2x(1-x)+m^2]}{3\Delta_R^3}\big]
	+\int_0^1 dx(1-x)^2\big[\frac{1}{\Delta_R}
	-\frac{q^2_{\bot}x(1-x)}{\Delta_R^2}\big]\Bigg\}\Bigg].\label{pipi0}
	\eea
	where $\Delta_R=m_p^2-q^2x(1-x)$.
	Similarly, $\Pi_{\pi^\pm}(q)^{PS}, \Pi_{\pi^0}(q)^{PV}$ and $\Pi_{\pi^\pm}(q)^{PV}$ can be derived likewise (for detailed discussion please refer to Ref.\cite{adhya17}).
	Finally, defining the effective pion masses by the positions of the pole of the propagator, 
	\be
	m^{*\,2}_{\pi} = m^2_{\pi} +{\rm Re}\,\Pi(m^{*\,2}_{\pi},{\bf q}=0,B),\label{effmass}
	\ee
	we calculate the variation of the mass with respect to the weak magnetic field.
	\section{Results}
	We present the variation of charged and uncharged pion masses with respect to the strength of the external magnetic field (according to the condition $(eB)<<m_{\pi}^2$ for the weak field regime). For our case, we show the results considering both the pseudo-scalar and pseudo-vector interaction of the $\pi-N$ Lagrangian in Figs.(1) and (2) respectively. We observed that the effective pion mass decreases for the PS interaction, whereas there is a slight increase for the PV interaction.
	\begin{figure}[h]
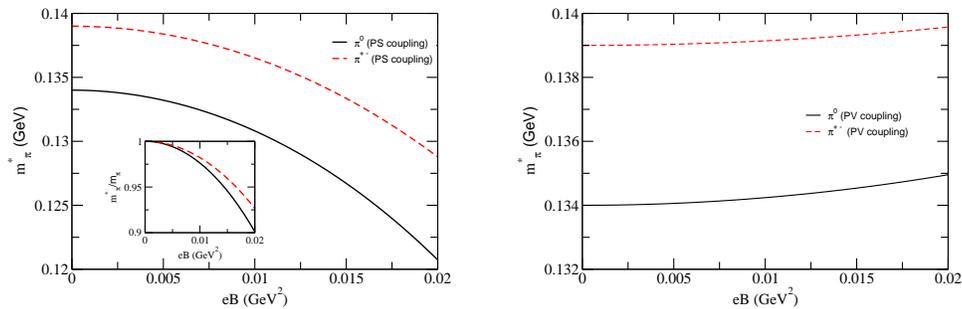

		\includegraphics[height=4cm, angle=0]{pi_PS_mass.eps}~~~~~~~~\includegraphics[height=4cm, angle=0]{pi_PV_mass.eps}
		\caption{ (Color online) Effective pion mass as a function of the magnetic field for PS and PV coupling.}\label{fig3}
	\end{figure}
	%
	%
	\vspace{-0.7cm}
	\section{Summary and discussions}
	The motivation of the present work has been to analyse the contribution of the weak magnetic field on the pion mass. We have introduced the weak magnetic field correction at one loop order through Schwinger's proper time method in the pion-nucleon Lagrangian. Our results show correction at $(eB)^2$ order for the field effects on the pion mass over the vacuum results. Finally, we have observed that the pion mass decreases for the PS coupling, whereas for the PV coupling, there is a marginal enhancement in mass for both the charged and uncharged pions \cite{adhya17}. The mass modification in medium due to magnetic field in realistic scenarios like neutron star or heavy ion collisions will be reported shortly.

	

\begin{thebibliography}{6}
		\vspace{-0.2cm}
		\bibitem{Tuchin:2013bda}
		K.~Tuchin,
		Phys.\ Rev.\ C {\bf 88} 024910 (2013).
		\bibitem{Ayala:2015lta}
		A.~Ayala, C.~A.~Dominguez, L.~A.~Hernandez, M.~Loewe and R.~Zamora,
		Phys.\ Rev.\ D {\bf 92} 096011 (2015).
		
		
		\bibitem{Colucci:2013zoa}
		G.~Colucci, E.~S.~Fraga and A.~Sedrakian,
		Phys.\ Lett.\ B {\bf 728} (2014) 19.
		\bibitem{Ghosh:2016evc}
		S.~Ghosh, A.~Mukherjee, M.~Mandal, S.~Sarkar and P.~Roy,
		Phys.\ Rev.\ D {\bf 94} (2016)  094043.
		
		\bibitem{schwinger51} J. Schwinger, Phys. Rev. {\bf82}, 664 (1951). 
		
		\bibitem{adhya17} S. P. Adhya, M. Mandal, S. Biswas and P. K. Roy,
		Phys.\ Rev.\ D {\bf 93} 074033 (2016).
		
		
		
	\end{thebibliography}
\end{document}